\documentclass[twocolumn,amsmath,amssymb,showpacs,superscriptaddress,reprint]{revtex4-2}
\pdfoutput=1
\usepackage[T1]{fontenc}
\usepackage{graphicx}
\makeatletter

\usepackage[english]{babel}
\usepackage[T1]{fontenc}
\usepackage[utf8]{inputenc}
\usepackage{braket}
\IfFileExists{lmodern.sty}{\usepackage{lmodern}}{}
\usepackage{color}
\definecolor{darkblue}{rgb}{0,0,0.6}
\definecolor{darkred}{rgb}{0.6,0,0}
\definecolor{darkgreen}{rgb}{0,0.6,0}
\usepackage{setspace}
\usepackage{environ}
\usepackage{wasysym}
\usepackage[colorlinks=true,urlcolor=darkblue,citecolor=darkblue,linkcolor=darkred,hyperindex=true,hyperfootnotes=false]{hyperref}
\usepackage{mathtools}
\usepackage{mathrsfs}
\newcommand{\setword}[2]{%
  \phantomsection
  #1\def\@currentlabel{\unexpanded{#1}}\label{#2}%
}

\begin{document}
%TC:ignore

\title {Unveiling the Connection: ER Bridges and EPR in Einstein's Research}
\author{\vspace*{-2mm}Galina Weinstein}
\affiliation{Reichman University, The Efi Arazi School of Computer Science, Herzliya; University of Haifa, The Department of Philosophy, Haifa, Israel.} 

\begin{abstract}
This paper explores the ER bridges theory and its relationship with quantum phenomena. An argument can be made that the ER bridges theory does not explicitly address quantum phenomena and implies that Einstein intended to differentiate between individual particles within the ER bridges theory and the systems involved in the EPR paradox. However, this paper contends that Einstein held a distinct viewpoint. He endeavored to elucidate quantum characteristics by modifying general relativity, aiming to incorporate concepts such as local realism, separability, causality, and determinism without relying on the principles of quantum mechanics. To achieve this, he proposed representing elementary particles using parallel ER bridges connecting two flat sheets.   
\end{abstract}

\maketitle

\section{Introduction}

This paper explores the interplay between the Einstein-Rosen (ER) bridges theory and the famous Einstein-Podolsky-Rosen (EPR) paradox.

In sections ~\ref{2} and ~\ref{3}, I intend to delve into the EPR thought experiment introduced by Albert Einstein, Boris Podolsky, and Nathan Rosen in 1935, examining its formulations and the paradox it poses.

The subsequent sections ~\ref{4} focus on Einstein's ER bridge theory and the challenges it confronts. Here, I analyze the problems addressed within this theory and how Einstein, Rosen, and Podolsky attempted to tackle them. 
I explore the motivation behind their pursuit of a modified general relativity framework, aiming to represent by bridges elementary particles and processes in which several elementary particles take part.

Lastly, in section ~\ref{6}, I delve into ER bridges and their relationship with entanglement. 
This section explores Leonard Susskind and Juan Maldacena's modern ER = EPR conjecture, which presents a contemporary perspective on the potential link between ER bridges and the EPR thought experiment. By examining this conjecture in section \ref{7}, I ponder the possibility of Einstein's concepts, ER bridges, and the EPR paradox being intertwined unexpectedly.

\section{EPR: elements of reality} \label{2}

In early March 1935, Einstein, Rosen, and Podolsky submitted the manuscript of the EPR paper, "Can Quantum-Mechanical Description of Physical Reality Be Considered Complete?", to \emph{Physical Review}, and it was published on May 15, 1935 \cite{Einstein:1935}

Einstein, Rosen, and Podolsky used the EPR thought experiment to demonstrate that the Heisenberg uncertainty principle fails if quantum mechanics is complete.

They first defined an "element of reality": If we can predict the value of a physical quantity without disturbing the measured system, then that quantity is an element of reality.
In the EPR thought experiment, the state of particle $A$ is dependent on the state of particle $B$.

\textcolor{darkred}{Let's start with the momentum of particle $B$.} 

The momentum of particle $A$ is measured and is found to have a value of $p$. We can infer that the momentum of particle $B$ is equal in magnitude but opposite in direction, $-p$, due to the conservation of linear momentum. 

This inference allows us to predict the value of particle $B$'s momentum without actually measuring it. 

Therefore, according to Einstein, Rosen, and Podolsky, the momentum of particle $B$ is an element of reality.

\textcolor{darkred}{Similarly, let's consider the position of particle $B$.} 

We measure the position of particle $A$ along the $x$-axis and find it at position $x_1$.
According to the "conservation of relative position" (a principle invented by Einstein), we can conclude that particle $B$'s position is related to $A$'s position by a fixed separation, denoted as $x_0$.
Specifically, particle $B$'s position can be expressed as $x_2 = x_1 + x_0$. 

Again, without directly measuring the position of particle $B$, we can predict its value based on the measurement of particle $A$. Thus, the position of particle $B$ is considered an element of reality.

This suggests that particle $B$'s momentum and position can be simultaneously determined, contrary to the uncertainty principle.

Indeed, Einstein, Rosen, and Podolsky argued that these elements of reality, namely the momentum and position of particle B, contradict the Heisenberg uncertainty principle. The uncertainty principle states that momentum and position cannot be precisely known simultaneously. The uncertainty principle represents the commutation relationship between the operators for momentum $P$ and position $Q$, showing that they do not commute.

Einstein, Rosen, and Podolsky show that if quantum mechanics is complete, momentum and position, with two non-commuting operators $P$ and $Q$, can have simultaneous reality. 

According to Einstein, Rosen, and Podolsky, since the two particles no longer interact at the time of measurement, no real change can occur in the second particle due to anything that may be done to the first particle ~\cite{Einstein:1935}. This is the principle of \textcolor{darkred}{local realism}: particles $A$ and $B$ should only be influenced by their local interactions. 

However, when particle $A$'s momentum is measured, particle $B$'s momentum is inferred instantaneously without disturbing particle $A$. This violates local realism.

Einstein, Rosen, and Podolsky thus conclude: \textcolor{darkblue}{quantum mechanics is incomplete.}

If we accept entanglement's non-local nature, i.e., adopt the idea that quantum mechanics is complete (and probabilistic), and renounce objective reality, there is no paradox. 

A realistic interpretation seems to suggest that a measurement of particle $A$ affects the properties of the other particle faster than the speed of light would allow. The crucial point here is that using quantum entanglement to transmit information faster than the speed of light is impossible. 
Any measurement performed on one particle will only provide unpredictable results. We can, therefore, not use the EPR pair to transmit information faster than the speed of light. Hence, causality is not violated because a measurement performed on particle $A$ collapses the quantum state of that particle, causing it to take on a definite value. 

\section{EPR: Separability} \label{3}

After the publication of the EPR paper, Einstein significantly sharpened the formulation of the EPR argument.
He argued against the notion that the measurement of particle $A$ can have an immediate and direct effect on the state of particle $B$. 

\textcolor{darkred}{Einstein formulated a "Separation Hypothesis."} 

The "Separation Hypothesis" implies that once particles $A$ and $B$ no longer interact, their subsequent states become independent. The real state of particle $B$ should not depend on the specific measurement of particle $A$.

The "Separation Hypothesis" challenges the idea that the measurement of particle $A$ could have an immediate and direct influence on the physical state of particle $B$.

Therefore, Einstein perceived the EPR thought experiment as highly problematic for the following reasons. If the physical state of particle $B$ were determined by the measurement of particle $A$, then there would be two different wave functions associated with the same physical state of particle $B$: 
One $\psi$ function would correspond to the measurement performed on $B$ itself (near $B$), and the other $\psi$ function would depend on the measurement carried out on particle $A$.

Einstein argues that a complete description of a physical state should be unambiguous, meaning there should be a unique and definitive description of the state. But if two different wave functions can describe the physical state of $B$, then the description of the state becomes ambiguous and incomplete \cite{Popper:1959}.

In his \emph{Autobiographical Notes}, Einstein presented another iteration of the EPR thought experiment. He contended that resolving the EPR paradox would necessitate invoking either a metaphorical "telepathy" (the instantaneous correlation between the states of the two entangled particles) or a breach of the separability principle — two concepts he found implausible. Einstein thus deduced that the EPR paradox was fraught with inconsistencies, further underscoring his belief that quantum mechanics was incomplete \cite{Einstein:1949}. 

\section{ER bridges: A modified general relativity theory} \label{4}

A few days before the EPR paper was published, Einstein and his assistant Rosen submitted the manuscript of the ER bridge paper, "The Particle Problem in the General Theory of Relativity," to \emph{Physical Review}, and it was published on July 1, 1935, ~\cite{Einstein2:1935}.

In the ER bridge paper, Einstein and Rosen tried to \textcolor{darkgreen}{exclude singularities from the field theory}. 
As a last resource, Einstein represented material particles as singularities in the field theory. But lo and behold, Einstein and Rosen found a way to describe elementary particles without relying on the presence of singularities in the field theory. 

Instead of singularities, they decided to introduce bridges. These bridges were intended to represent particles. Using bridges instead of singularities, Einstein and Rosen aimed to develop a model free from the mathematical and conceptual difficulties associated with singularities. This idea of smoothed-out structures is somewhat aligned with modern theories like string theory, where particles are not point-like but are instead extended objects, namely strings, that vibrate at different frequencies.

Einstein and Rosen first modified the field equations to obtain a singularity-free solution. 

Particles are no longer represented by singularities in the field but by a topological structure; a bridge models them.

First, Einstein and Rosen tried to replace an elementary particle having mass but no charge with a topological structure, an ER bridge of finite length in four-dimensional spacetime. The spatially finite bridge was identified as the neutron, an elementary particle with mass but no electric charge. 

\begin{figure}
\includegraphics[width=1\columnwidth]{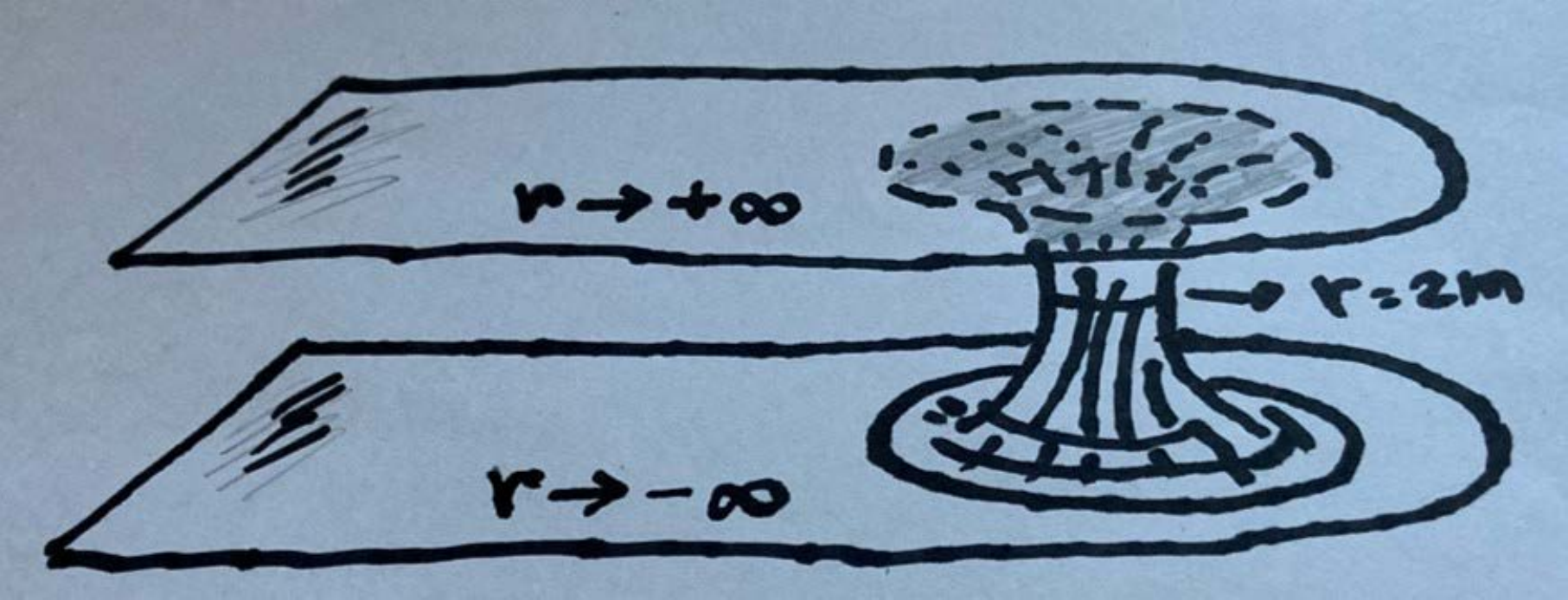}
\vspace*{-6mm}
\caption{Embedding diagram of an Einstein-Rosen bridge. According to Einstein and Rosen’s modified general relativity, four-dimensional spacetime is described by two congruent sheets connected by spatially finite bridges. The spatially finite bridges are elementary particles. Two types of bridges connect the two identical sheets; the first is the opposite of the second: the first type provides an alternative (mathematical) representation of a neutron with mass but no electric charge, while the second has charge but no mass.\label{Figure_1}}
\vspace*{-5mm}
\end{figure}

Einstein and Rosen then tried to replace an elementary charged particle with a spatially finite bridge. To obtain a bridge representing a charged particle, they considered the negative of the Maxwell stress-energy tensor in a vacuum and set the mass equal to zero. They obtained a solution free from singularities for all finite points in the space of two sheets, and a bridge between the two sheets again represented the charged particle. The ER bridge represented an elementary electric particle with electric charge but without mass (and negative energy density) (see figure ~\ref{Figure_1}). 

I will refrain from delving into historical intricacies and briefly recap the key details of Karl Schwarzschild's accomplishments. Schwarzschild discovered a solution to Einstein's field equations in general relativity, describing the gravitational field around a spherically symmetric mass. He found that for a collapsing sphere, there is a critical radius (called the Schwarzschild radius) where the field equations fail to provide a meaningful description. Beyond this radius, known as the "Schwarzschild Singularity," the sphere is thought to continue collapsing.

Einstein believed that discussing what happens beyond the Schwarzschild Singularity was meaningless because the field equations did not provide physical meaning for that region. Instead, he considered the Schwarzschild radius $r = 2m$ as the limiting point rather than $r = 0$, and therefore he did not explicitly mention "black holes" in his discussions.

Einstein and Rosen introduced a coordinate transformation to eliminate the region containing the Schwarzschild singularity at $r = 2m$. They introduced a new variable, $u^2 = r - 2m$, which allowed them to express the Schwarzschild solution as a regular solution without any singularities for all finite points and values of $u$.
The new solution represented physical space as a mathematical construct consisting of two congruent identical flat sheets joined by a bridge at $r = 2m$ (see figure ~\ref{Figure_1}).  

However, the ER bridge theory was not without its challenges, and soon enough, issues began to arise, casting a shadow over its initial promise. 

\begin{figure}
\includegraphics[width=1\columnwidth]{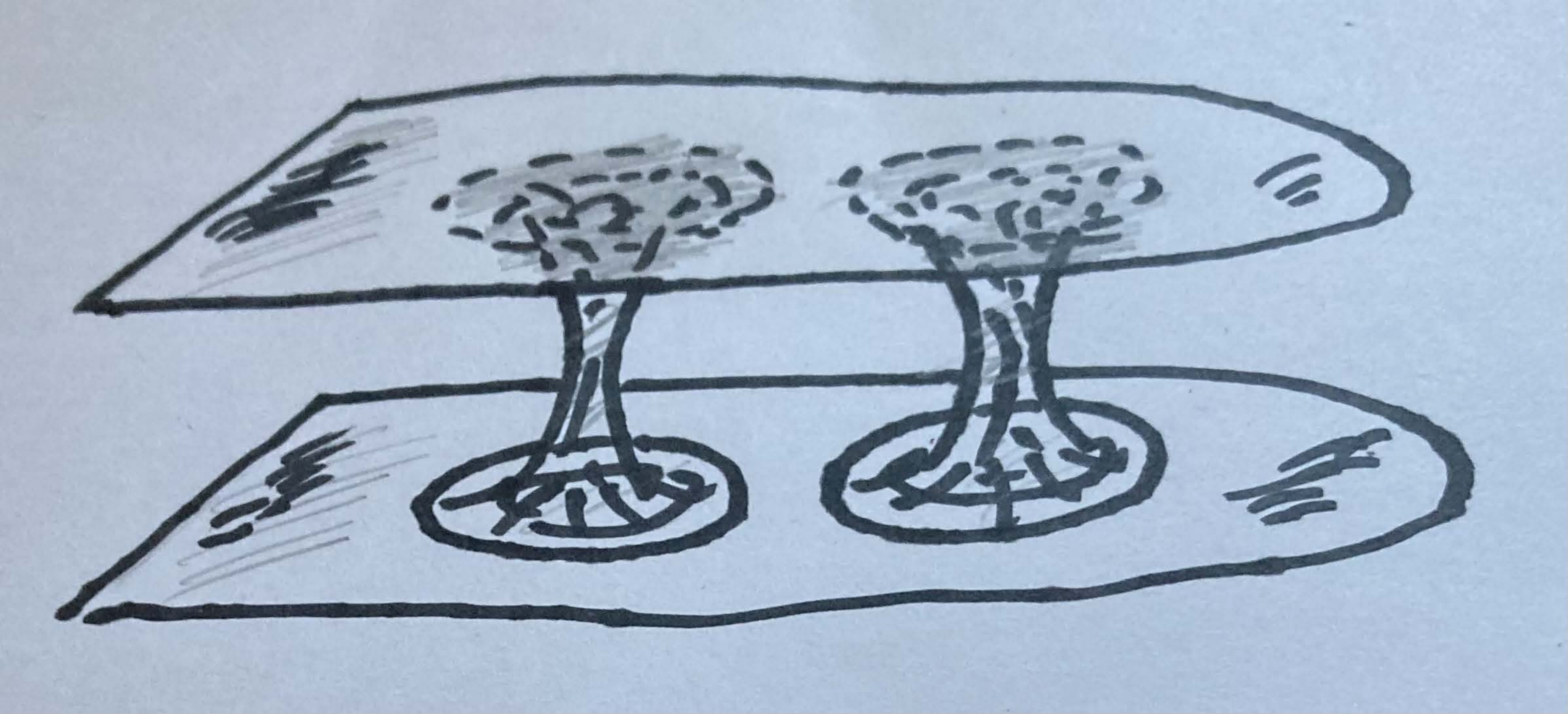}
\vspace*{-6mm}
\caption{An Einstein-Rosen bridge pair. Einstein and Rosen: “One is therefore led, according to this theory, to consider the electron or proton as a two-bridge problem.” \cite{Einstein2:1935}. The two bridges for mass and charge are distinct entities connected by the same flat sheets.\label{Figure_2}}
\vspace*{-5mm}
\end{figure}

The electron, proton, and positron (anti-electron having the same mass and opposite charge to an electron) all have mass and electric charges. 
Einstein and Rosen concluded that each particle would be represented by two bridges between the congruent, identical flat sheets (see figure ~\ref{Figure_2}). 

They further expected that processes in which several elementary particles take part correspond to regular solutions of the field equations with several bridges connected to the two identical flat sheets (\cite{Einstein2:1935}).

In this situation, the two flat sheets would become somewhat congested with parallel ER bridges, creating overcrowding.

\textcolor{darkred}{But Einstein and Rosen encountered difficulties mathematically constructing a solution to the field equations of general relativity that accommodated multiple ER bridges.} They were uncertain about regular solutions with more than one bridge \cite{Einstein2:1935}.

In a letter to his closest friend Michele Besso, \textcolor{darkred}{Einstein mentioned his collaboration with Podolsky}, referring to him as a “young colleague” and a “Russian Jew,” and \textcolor{darkred}{expressed their relentless struggle in addressing the many-body problem in the ER bridge theory} (Einstein to Besso, February 16, 1936, in \cite{Speziali:1972}); \cite{Weinstein:2022}. 

\textcolor{darkblue}{Einstein made many fruitless efforts to find a field representation of matter free from singularities based on a unified field theory program. Einstein and his assistants attempted to find a precise solution to the gravitational field equations by considering the electromagnetic field and avoiding singularities. They based their investigation on ER bridges but ultimately concluded that no such solution exists} (\cite{Einstein2:1938}).

\section{ER = EPR and AdS/CFT} \label{6}

In this section, I will delve into the contemporary facets of the ER = EPR conjecture and the AdS = CFT theory.

Suppose we create maximally entangled pairs of particles, specifically EPR pairs of particles. We can describe this non-local correlation from the point of view of a conformal field theory (CFT).

According to the ER = EPR conjecture, if we describe the entangled particles in an Anti-de Sitter (AdS) space, an ER bridge provides a geometric representation of the entanglement between the EPR particles. 
On the other hand, an ER bridge between two black holes implies that they are entangled; entanglement between two black holes implies that an ER bridge connects them. 

\begin{figure}
\includegraphics[width=1\columnwidth]{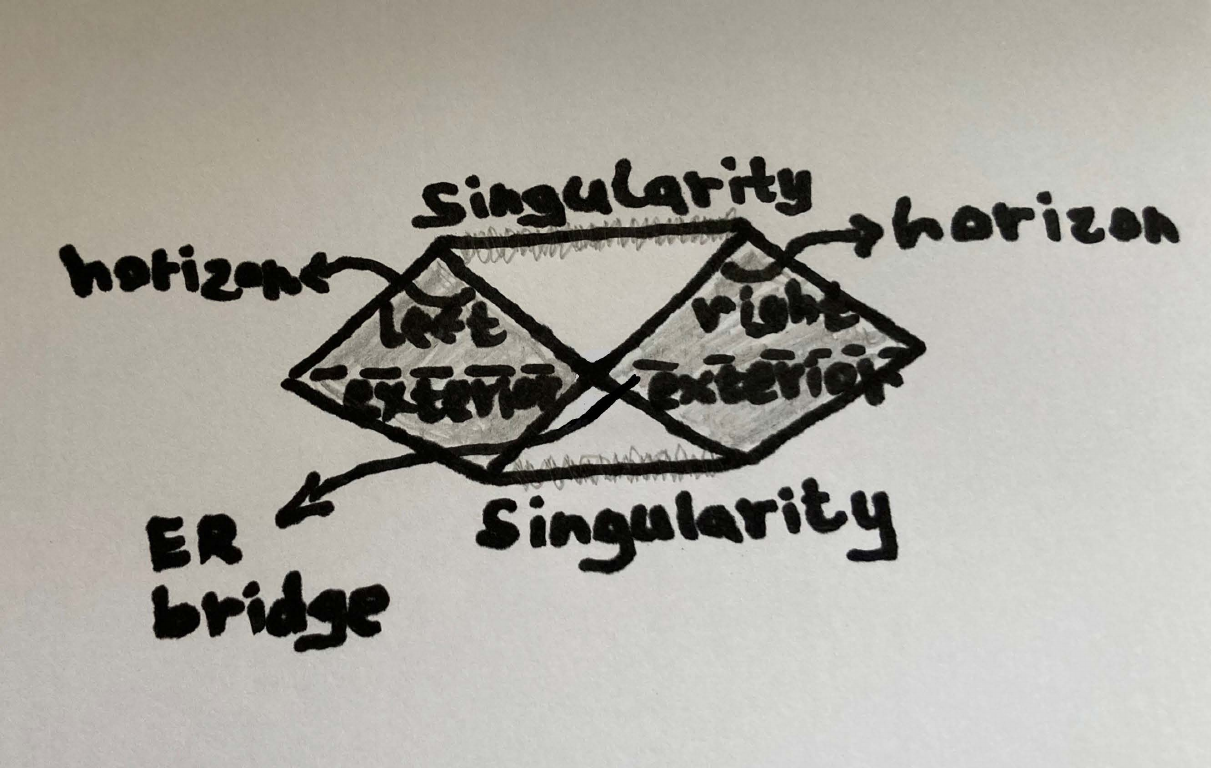}
\vspace*{-6mm}
\caption{The Penrose diagram of an AdS asymptotically two-sided eternal black hole. This is inspired by a drawing in \cite{Maldacena2:2013}. According to Susskind, "we examine the Penrose diagram of an AdS 'eternal' black hole or what is really two entangled black holes connected by an Einstein-Rosen bridge" \cite{Susskind:2020}. \label{Figure_3}}
\vspace*{-5mm}
\end{figure}

When we describe two distant black holes as entangled in the gravity picture, it does not imply that their quantum states are correlated. Instead, this statement suggests that an ER bridge (or a wormhole) connects the two black holes through their interior regions \cite{Maldacena:2013}:
Alternatively, one can say that an ER bridge connects the two exterior regions of an eternal black hole. 
The entanglement between two CFTs can be dual to the ER bridge connecting the exterior regions of the eternal black hole (see figure ~\ref{Figure_3}). 

Another perspective is that the thermofield double (TFD) state corresponds to a dual representation of two black holes. An ER bridge comes into play to connect the interior regions of two black holes. Each of the CFTs is linked to one of the black holes.
To justify the claim that the eternal black hole is dual to the TFD state, we apply the AdS/CFT dictionary.

In their paper, “A holographic wormhole traversed in a quantum computer,” Adam Brown and Susskind highlight that "the idea of a "wormhole dates back to 1935" when Einstein and Rosen \cite{Einstein2:1935} "studied black holes in the context of Einstein's general relativity. [...] In the same year, Einstein and Rosen wrote another paper \cite{Einstein:1935}, this time in collaboration with Boris Podolsky. This trio's paper examined quantum mechanics (without gravity) and identified the phenomenon now known as quantum entanglement [...] At the time, these two ideas — wormholes and entanglement — were considered to be entirely separate" (\cite{Brown:2022}). 

Brown and Susskind delve into an intriguing aspect of the ER = EPR conjecture, which pertains to its connection to the reversibility of quantum mechanics. Susskind, Maldacena, and other researchers in the AdS/CFT field have been engaged in discussions regarding the concepts of scrambling and unscrambling. This concept is utilized in various quantum information protocols, Scrambling is when the information associated with a particle entering a wormhole becomes highly disordered. However, this information is not truly lost within the wormhole ~\cite{Brown:2019}, ~\cite{Brown:2022}. 
The principle that allows for the retrieval of the original quantum state is the inherent reversibility of quantum mechanics. Due to this reversibility, it is possible to apply a sequence of operations - known as state unscrambling - that are the inverse of the original scrambling operations. By doing so, one can retrieve the original quantum state of the particle.

Obviously, reversibility is a feature associated with closed quantum systems, where no information is exchanged with an external environment. In open quantum systems, environmental interactions can lead to irreversible processes like decoherence, making it difficult or even impossible to unscramble information perfectly. However, a quantum circuit can be constructed to model the dynamics of scrambling and unscrambling. The system is considered closed in the idealized quantum circuit model, meaning it doesn't interact with an external environment. Theoretically, we can always find a set of gates that will act as the inverse of the scrambling operations. 

\begin{figure}
\includegraphics[width=1\columnwidth]{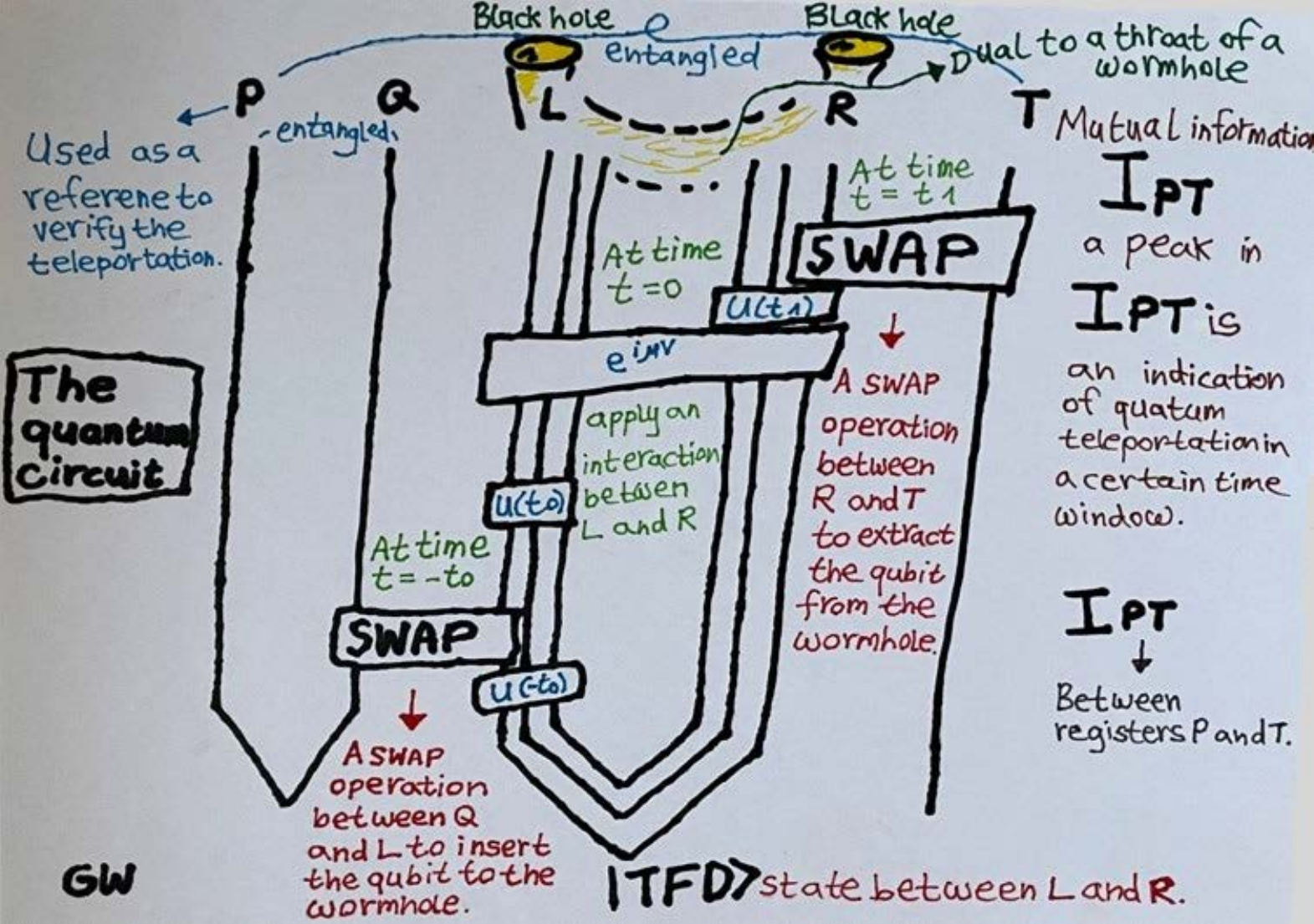}
\vspace*{-6mm}
\caption{The quantum circuit of the traversable wormhole is composed of the following elements: researchers prepared the thermofield double (TFD) state $\ket{\text{TFD}}$, which consists of the eigenstates of the left and right quantum systems. They then created a maximally entangled state between registers $P$ and $Q$. At the time $t = -t_0$, they applied a SWAP operation between registers $Q$ and $L$ to insert the qubit into the wormhole, creating an entangled connection between them. In the quantum picture, the SWAP between $Q$ and $L$ replaces the states of the qubits in those registers. As a result, the qubit in $R$ becomes entangled with the new state in $L$ that originated from $Q$. The gravity picture describes this as "inserting the qubit into the wormhole."
At the time $t = 0$, an interaction is applied between the left and right quantum systems, carried out between register $R$ and register $L$. Finally, at the time $t = t_1$, another SWAP operation is applied between registers $R$ and $T$, effectively exchanging their quantum states. This SWAP can be seen as "pulling" the quantum information from one side of the wormhole $L$ to the other $R$, much like teleportation. In this way, the qubit's state has been moved "across the wormhole" without the qubit itself physically traversing it \cite{Weinstein2:2022}.\label{Figure_5}}
\vspace*{-5mm}
\end{figure}

Researchers have explored the idea that the highly entangled, scrambled states could facilitate a sort of quantum teleportation through the ER bridge. Information is scrambled on the left side of the wormhole. Due to the intrinsic connection between the two sides of the wormhole - left and right - this information is then unscrambled and emerges on the right side (see figure ~\ref{Figure_5}).  

\section{ER bridges and entanglement} \label{7}

In Maldacena's paper, "Black Holes, Wormholes and the Secretes of Quantum Spacetime," the assessment of Einstein's work, along with that of his colleagues, dating back to 1935, closely aligns with Brown and Susskind's perspective \cite{Maldacena:2016}: 

\begin{quote}
\textcolor{darkblue}{"Interestingly both quantum entanglement and wormholes date back to two articles written by Albert Einstein and his collaborators in 1935. On the surface, the papers seem to deal with very different phenomena, and Einstein probably never suspected that there could be a connection between them".}    
\end{quote}

Like Susskind, Maldacena ascribes to Einstein certain aspects of contemporary thought that diverge significantly from Einstein's original thinking and motivations.

Unlike the ER = EPR conjecture, Einstein and Rosen's original bridge represents an elementary particle. \textcolor{darkgreen}{It does not inherently imply any correlation or entanglement between the elementary particles.} \textcolor{darkblue}{The ER bridges are separate structures, each representing one particle and two separate bridges representing the electron/proton. Thus, the particles the ER bridges represent behave according to their individual properties.} This is most reasonable because Einstein adhered to the belief of local realism.

Another crucial aspect to take into consideration is the
following. \textcolor{darkred}{From Einstein's standpoint, duality does not revolve around the relationship between entanglement and ER bridges. Instead, the duality was between gravity and elementary particles. The purpose of the bridges is to reconcile the atomistic theory of matter with the principles of general relativity, effectively eliminating this duality.} 

\begin{figure}
\includegraphics[width=1\columnwidth]{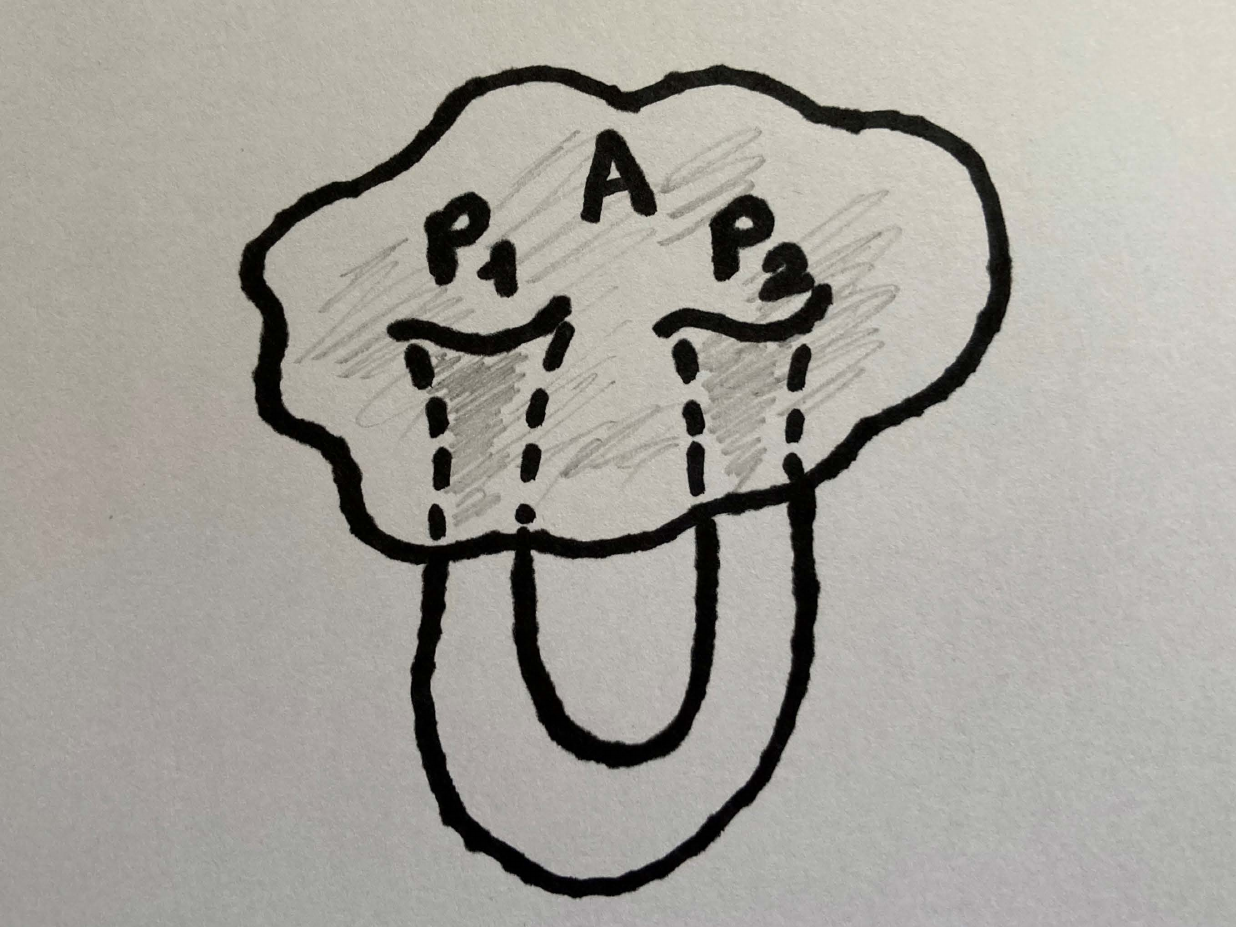}
\vspace*{-6mm}
\caption{The two mouths $P_1$ and $P_2$ of the wormhole are connected to spacetime $A$ \cite{Wheeler:1957}.\label{Figure_4}}
\vspace*{-5mm}
\end{figure}

\textcolor{darkblue}{In 1935, Einstein did not consider the ER bridge a wormhole. Nor did he associate it with two black holes or mention “black holes” in his discussions.} The modern interpretation of the ER bridge as a wormhole is attributed to John Archibald Wheeler, who introduced the concept of wormholes after Einstein’s death. In a paper presented in 1957 at a conference in Chapel Hill, North Carolina, Wheeler included a drawing of a two-mouth wormhole (see figure ~\ref{Figure_4}) and wrote that in a two-dimensional space, a "wormhole" allows for the connection of two regions, $P_1$ and $P_2$, enabling an ant traveling from $P_1$ to instantly appear at $P_2$ without leaving the initial two-dimensional surface \cite{Wheeler:1957}. 

The description of the wormhole above as an ER bridge lacks physical validity and historical accuracy. However, it is plausible to suggest that Wheeler drew inspiration from Einstein's 1935 topological framework. 

It could be argued, with some justification, that the ER bridges theory does not explicitly tackle quantum phenomena. Moreover, in the context of the EPR argument, Einstein consistently referred to systems $A$ and $B$ rather than individual particles $A$ and $B$ (see, for instance, \cite{Einstein:1949} and \cite{Einstein:1936}). One could argue convincingly that this suggests Einstein intended to clearly distinguish between the individual particles in his ER bridge theory and the systems involved in the EPR paradox. 

However, I suggest that Einstein held a different perspective. \textcolor{darkred}{Einstein expressed dissatisfaction with the incompleteness of general relativity.} He argued that one of the limitations of general relativity is its incompleteness as a field theory, as it necessitates the introduction of an independent postulate concerning the motion of particles defined by the geodesic equation. On the other hand, a complete field theory exclusively describes phenomena using the language of fields, and material particles are solutions of the field equations with no singularities at all \cite{Einstein2:1935}. 

\textcolor{darkblue}{Einstein aimed to elucidate the quantum characteristics of fundamental particles by revising general relativity. His objective was to reintroduce the notions of local realism, separability, causality, and determinism into this modified framework of general relativity and field theory without relying on the principles of the newly developed quantum mechanics, which entailed randomness and probabilistic outcomes. He proposed the idea of representing elementary particles using ER bridges but ultimately faced challenges in his endeavors with Podolsky.}  

Einstein ardently pursued a unified field theory. However, during his time, the advanced tools and concepts needed to discern a profound link between quantum entanglement, as illustrated by the EPR thought experiment, and the spacetime topology, as later explored in the ER = EPR correspondence, were still in their infancy. However, Einstein's ideas on the EPR paradox and ER bridges have been instrumental in shaping the contemporary conjecture ER = EPR and theory AdS/CFT. Even though Einstein did not directly conceive these modern concepts, his work provided the essential groundwork from which they emerged.

\begin{acknowledgments}

\noindent This work is supported by ERC advanced grant number 834735. 

\end{acknowledgments}

\bibliographystyle{apsrev4-2}
\bibliography{references}

\end{document}